# On the complexity/criticality of Jamming during the discharge of granular matter from a silo"

P. Evesque

Lab MSSMat, UMR 8579 CNRS, Ecole Centrale Paris
92295 CHATENAY-MALABRY, France, e-mail: pierre.evesque@ecp.fr

**Abstract:**

*This paper is aimed at pursuing a recent discussion about the comparison between Self-Organised Criticality, the jamming process and the percolation theory in the problem of a silo discharge [I. Zuriguel, A. Garcimartin, D. Maza, L.A.Pugnaloni, J.M.Pastor, " Jamming during the discharge of granular matter from a silo ", Phys.Rev.E 71, 051303 (2005)]. Statistics of blocking a silo is investigated from different models: percolation, self organised criticality..*

**Pacs # : 5.40 ; 45.70 ; 62.20 ; 83.70.Fn**

It is known that the flow of grains from a silo stops spontaneously when the aperture diameter D of the outlet is too small compared to the grain diameter d; it means that this flow needs to be restarted "periodically" for the silo continues emptying. This phenomenon has been known from a while, but it has been accurately and extensively studied recently [1] under well defined condition when the restarting process is some air blown in the outlet. It has been confirmed that the right parameter that controls blocking is D/d=R; it has also been found that spontaneous blocking occurs for $R<R_c=4.94\pm0.03$ for spheres under the experimental conditions, and that $R_c$ depends slightly on the grain geometry. If one defines the size s of an event as the number of grains that fall from the outlet between two successive stops of flow, then statistics $n_R(s)$ can be drawn for different R. The distribution $n_R(s)$ has been found to become rather broad just below $R_c$; also the distribution of events are characterised by two sizes, the first one is called the mode $s^*$ and the second one, which is much larger, is the typical (or average) size $<s>$: $s^*$ is the size of the event below which the avalanche remembers that it has started recently; and $n_R(s)$ increases till $s^*$; then $n_R(s)$ decreases continuously above $s^*$. So, it was found that the stopping process looks independent of time above $s^*$, and that the large events (flow $s>s^*$) are characterised by a statistics $n_R(s)$ of flow size s which exhibits an exponential tail $n_R(s)= n_{Ro} \exp(-s/<s>)$ with an extinction size $<s>$ which depends on R and varies as $(9900\pm100)/(R-R_c)^{\gamma}$, with $\gamma=6.9\pm0.2$; on the contrary, for small event $(s<s^*<60)$ the statistics follows a power law, *i.e.* $n_R(s)= (s/s^*)^{\kappa}$ , with $\kappa=2$-(Fig.4 of [1] to 3-4 [1]. This can be summed up as follows:

∃ 2 sizes : the Mode $s^*$ and average size $<s>$ ($>s^*$)

$$n_R(s)= n_{Ro} \exp(-s/<s>) \qquad \text{for } s > s^* \qquad (1.a)$$

$$n_R(s)= (s/s^*)^K \qquad \text{for } s < s^* \,(<60) \qquad (1.b)$$

So when $R<R_c$ the flow occurs by a series of discrete events, which one can call "avalanche".

In the recent literature, such a stochastic flow is often analysed in terms of self organized criticality (SOC) [2]. So one is led to compare or interpret the experimental data within this scheme (SOC). In [1] also, the size distribution s of events that occur between two stops is compared to the distribution of finite clusters in a percolation problem in 1d. In particular, ref. [1] has proposed to associate $R_c$ with the threshold $p_c$ of a percolation problem [3] and to associate $(R-R_c)$ to the control parameter.

This paper is aimed at pursuing the discussion about the comparison between SOC, the jamming process and the percolation theory in the problem of a silo discharge [1].

## 1. Self-Organised Criticality (SOC):

In Self-Organised Criticality, first introduced by Bak-Tang & Wiesenfeld [2], the flow of sand on a free surface was supposed to occur without generating a typical length scale so that a distribution of avalanche size was expected to obey a power law distribution, *i.e.* $n_R(s)= s^{-\tau}$. This result was thought to happen because avalanches were viewed as a surface flow which occurs at a given inclination and stops at the same inclination, with definite rules which drive the system to and force it to remain at this inclination. In practice, it has been found that there were two different angles (a starting $\theta_{start}$ and a stopping angle $\theta_{stop}$) leading to define (i) a typical avalanche size $S^*$ and (ii) that this avalanche size scales as the pile volume $S^*=(L/d)^3(\theta_{start}-\theta_{stop})$ (and not as the pile surface) so that the real avalanche size is always much larger that the one predicted by SOC and scales as a volume. (It is only in the case of very small pile for which $L(\theta_{start}-\theta_{stop})<d$ that one observes the critical behaviour) [4].

Indeed this approach is not valid here in the case of the flow from a silo, for which two typical sizes $s^*$ and $<s>$ have been defined and found in [1]. Furthermore, SOC system is supposed to adjust itself so that its working point becomes a critical point (in the sense of critical phenomena theory and of bifurcation theory). This is not true obviously for the flow from a silo, because it exists a typical diameter size $D_c$ below (above) which the flow does (does not) stop.

In turn as the flow from a blocked silo can be restarted at will by some local perturbation (at least in experiment reported in [1]), the mechanism does not imply as much "hysteresis" as in the avalanche flow [4]. This results probably from a difference in set-up: in silo flow, the flow is controlled by what occurs at the outlet, so that it concerns a finite volume, while in a surface-avalanche flow the flow is controlled by what occurs at the whole surface.

## 2. Jamming:

The diameter $D_c$, or the diameter ratio $R_c$, can look as a critical value of the parameter at which a jamming/unjamming bifurcation occurs. It is then important to characterise the behaviour of this jamming transition. It has been often stated that jamming in granular material results from a complex mechanical process due to interlocking of grains which ensures force chains propagation on large distances, so that behaviour is complex and scales in a non regular manner with the size of the system [5]. Here we will see that it is not true since maximum size of events below the mode remains small (<50grains) with typical correlation length $\xi/d$ ratio <5. We will associate $\xi$ to the maximum size $s^*$ of the mode. No correlation at larger time/size scale exists for larger events most likely, because these events obey Poisson distribution. Does this strengthen the analysis of jamming as the domain of validity of quasi-static soil mechanics [6a]?

## 3. Percolation:

Let us now recall some well known results on percolation theory [3]. Percolation is used to describe the flow through a porous medium whose holes are closed (with probability 1-p) and open (with probability p). One finds in this case that some of the holes are only connected to a finite number of holes, forming a finite cluster, but that it exists also a threshold $p_c$ for and above which an infinite cluster exists through which the flow can propagate over an infinite distance. It is found that $p_c$ depends on (i) the nature of the bonds between holes, (ii) on the number of neighbouring holes and (iii) on the dimensionality of the space (2d, 3d,…). It is found also that the system exhibits a complete scaling invariance just at $p_c$ which makes large finite clusters looking as a part of the infinite cluster…. , so that cluster distribution, cluster shape ,… obey scaling laws which are governed by critical exponents. It is also found that these critical exponents do not depend on the local bond geometry or local lattice structure, but only on the dimensionality of the space. At last, near the threshold $p_c$, it exists a range of ($p-p_c$) for which similar scaling works for the finite clusters which are small enough, so that it exists a typical maximum size of finite cluster $S^*$ ; and $S^*$ depends on ($p-p_c$) through a critical exponent.

So, as recalled in last paragraph, percolation concerns a geometrical problem of flow through a heterogeneous medium, which generates complexity of cluster topology at large length scale. Scaling of the complexity does not depend on the details of the local interaction, but depends strongly on the dimensionality of the lattice space itself (1d, 2d, 3d, …). It spans to all distances smaller than the maximum size of the finite clusters (which remains finite for $p-p_c \neq 0$, but becomes infinite at $p=p_c$). In particular, cluster ramifications are important in 2d, 3d,..., leading to fractal structure network; mean field theory applies only above 6d. On the contrary, complexity is very limited in 1d percolation problem because the topology of finite clusters is always linear so that cluster boundary is finite and constant.

In other words, near the threshold $p_c$, the distribution n(s) of finite clusters follows the scaling

$$n(s) \propto s^{-\tau} f[(p-p_c)s^{\sigma}] \qquad (2)$$

where $\tau$ and $\sigma$ are critical exponents; noting $z=(p-p_c)s^\sigma$ one gets also that f(z) exhibits an exponential tail: $f(z) \propto \exp(-z/z_c)$ at large z. $\tau$ and $\sigma$ are related to the other classic critical exponents: $\tau=(2\nu d-\beta)/(\nu d-\beta)$, $\sigma=1/(\nu d-\beta)$. One notes also that $s^{-\tau}$ is a decreasing function of s; however moments of the distribution which are of order higher than $\tau-1$ diverge at $p_c$. The size distribution (Eq. (2)) governs the physics near the percolation threshold, for instance, the anomalous distribution of small clusters $n(s) \propto s^{-\tau}$ is controlled by the first part of the right hand side of Eq. (2), while the function f, which decays exponentially, defines the maximum size $S^*$ of these finite clusters with anomalous distribution: $S^* \propto (p-p_c)^{-1/\sigma}$. Parameters of physical interest are often related to the different momenta , $[\Sigma_s s^k n(s)]$, of this distribution. It can be shown that their singular part depend on $(p-p_c)^q$, via the exponent q which is called a critical exponent. This exponent q depends on $\tau$ and $\sigma$: using Eq. (2), one can find the exponent $\alpha$ from its definition as a critical exponent:

$$[\Sigma_s n(s)]_{sing} \propto (p-p_c)^{2-\alpha} \qquad (3)$$

So, applying Eq. (2), Eq. (3) becomes $[\Sigma_s s^{-\tau} f][(p-p_c)s^\sigma]$, which gives after a change of variable $z=(p-p_c)s^\sigma$ and after identification $2-\alpha=(\tau-1)/\sigma$. Similar procedure can be done with exponent $\beta$, $\gamma$,…and one gets the following relations $\beta=(\tau-2)/\sigma$, $-\gamma=(\tau-3)/\sigma$. Similarly, the correlation length $\xi$ is related to $(p-p_c)$ via the critical exponent $\nu$: $\xi \propto (p-p_c)^{-\nu}$ and to the maximum size of finite cluster $S^* \propto \xi^D$ through the fractal dimension $D_f$ of the cluster, $D_f=d_o-\beta/\nu$, where $d_o$ is the space dimensionality ($d_o=2$ or 3). In classic terminology [3], one has also the relations $\{\alpha=2-\nu d_o;\ \gamma=(\nu d_o-2\beta);\ \delta=(\nu d_o/\beta-1)\}$.

In the case of a percolation problem in one dimension, the nature of the problem simplifies into a Poisson problem [3], due to the constrains imposed by the 1d geometry : in particular one finds the fractal dimension $D_f=1$ in this case, and the threshold is $p_c=1$. One gets also $\xi=-1/\ln(p) \approx 1/(p_c-p)_{\text{when } p \to p_c=1}$, $\nu=1$, $\alpha=1$, $\beta=0$, $\tau=2$, $\sigma=1$. It means that all "critical" exponents are integers and that the geometry of cluster is linear. In other words the critical behaviour reduces to classic behaviour, which is not "critical".

## 4. Discussion:

After these recalls, we can proceed to the comparison of what was found in [1] to what is expected for a system that obeys SOC approach, and to give some conclusion about the "jamming" transition and its nature.

*Firstly,* it has been found in [1] that one can define a probability p that a grain passing through the aperture does not block the aperture of radius R, that the distribution of large events obeys a "percolation modelling in 1d". It has been found also in [1] that the threshold $p_c=1$ of this percolation corresponds to a single value of the aperture $R=R_c$. It means that the system does not converge towards a SOC behaviour, but that it is kept spontaneously at a given distance to it, and that this distance $(p-p_c)$ is fixed by R (or $R-R_c$) so that the typical avalanche size $<s>$ varies as $A(R_c-R)^{-\gamma}$,

$$<s> \propto A(R_c-R)^{-\gamma} \propto B(p_c-p)^{-1} \qquad (4)$$

with $\gamma=6.9\pm0.2$ In other word, there is a fix transition point at $R_c =4.94\pm0.03$ that correspond to $p_c=1$ and a one to one correspondence between p and R.

*Secondly,* we use what is recalled in §-3; so if one uses a 1-d percolation modelling to analyse the size of the discrete series of continuous flows, it means that each cluster of te percolation is one event (continuous flow of size s). It means also that the distribution of cluster size obeys a Poisson distribution characterised by (i) a fractal dimension $D_f =1$, (ii) a critical threshold which is $p_c=1$, (iii) that the correlation length $\xi=-1/\ln(p) \approx 1/(p_c-p)_{\text{when } p\to p_c=1}$, (iv) that other critical exponents are $\nu=1$, $\alpha=1$, $\beta=0$, $\tau=2$, $\sigma=1$. It means then that all "critical" exponents are integers and that the geometry of cluster is linear. In other words the critical behaviour reduces to classic behaviour, which is not "critical".

Also, since 1d percolation predicts $<s> \propto 1/(p-p_c)$ and that experiments find Eq.(4), this fixes the relation between $(p_c-p)$ and $(R_c-R)$.

*Thirdly,* we have now to discuss whether the flow can be mapped over a percolation modelling. What is found experimentally is that any grain can block the outlet with probability (1-p) when passing through the outlet. It means simply that the outlet plays the role of a single hole which can be closed or not by a grain. Obviously, this is not equivalent to the problem of the flow through a continuous porous medium with a geometrical distribution of connected/unconnected holes. In the present case, the only dimension which allows a "geometrical" description of the clusters is the time (and not the space). So the problem is much better stated in terms of a Poisson process with no correlation in time, rather than in terms of a percolation problem or in terms of a phase transformation, which impose cooperative processes. Here the problem is just the problem of a tap which is open or closed alternatively with probability p and (1-p). it is difficult to consider the tap as a porous medium.

The problem of the mechanical coherence of the stopper in the tap is not needed to be discussed, because no measurement is done on how difficult breaking the stopper coherence is. It may be tiny, *i.e.* near a mechanical percolation threshold, with a backbone having a fractal structure,…, or much stronger, that does not play any role in the experimental result: What we know is just that the jet of air is sufficient to break it. So we do not think that percolation has to be used as a term characterising the stopper itself.

In other words, one can see the flow generating an arch of free fall at the outlet, and any grain, when it passes through the outlet, is triggering the stabilisation (blocking) of the arch with probability (1-p). It is worth recalling that the arch of free fall is a hydraulic jump we describe a little more in point (vi) below; here what we consider is that this dynamic structure is stabilised (transformed) in static vault with probability (1-p) when a grain crosses the arch.

*Fourthly,* to settle the problem as a cooperative process, one has to introduce what causes the blockage of the outlet. Indeed, it is a cooperative process which concerns N grains at least. But as nothing is known about the true configuration needed for blockage and the whole distribution of grain configurations, things happen as if there were no time correlation. To study this cooperative process, one should investigate correlations of flow fluctuations at short time: be Q(t) the flow of grains from the outlet, <Q> its mean and $N_c$ the typical size needed to make a tap, one should look to the evolution of the correlation function <Q(t)Q(t-$\tau$)> for 0<$\tau$ < 10 $N_c$/<Q> and its variation when a blockage is reached.

Anyhow, let us assume that the formation of a static vault requires some geometrical structure $V_{ault}$ made by $N_v$ grains. The grain configuration at the outlet changes continuously with time so that the probability of generating the blocking structure happens uniformly with time when flow occurs (if flow remains uniform in time). If flow does not remain uniform in time, the change of configuration at the outlet depends linearly on the number which flows through the outlet most likely, so that can assume that the probability of static Vault formation is proportional to the number of grains having flowed through the aperture, which leads to define a Poisson process which depends on s.

*Fifthly,* experiments show that there are two typical flow size $s^*$ and <s>. In percolation there exists a single size <s>, with a number of clusters n(s) scaling as n(s) $\propto s^{-\tau}$ for s < <s>. Can one consider the power law variation of n(s) $\propto s^{\theta}$ before s=$s^*$ as a critical behaviour? **No** because the experimental exponent is found to be positive instead of negative (-$\tau$); a positive $\theta$ indicates a divergence at large s; in other words, it would tell that the larger the events the more numerous. In this case the physics will be controlled completely by the larger events. But as $s^*$ remains small, and does not correspond to <s>, the growth stops very early contrarily to what is expected in percolation theory.

*Sixthly,* the size of the mode $s^*$ in [1] has been attributed to the existence of a transient regime. In fact the restart of the hourglass flow after a stoppage requires some time (and some amount of grains) before a permanent regime occurs. Does this time corresponds to $s^*$? This is what we want to study in this paragraph. It is often assumed that the permanent flow from the outlet of a hourglass is limited by some a hydraulic jump at the outlet; this one is generated by the effect of the gravity acceleration combined with the convergence of the flow lines: indeed, be z the distance to converging point of the flow lines, the flow of granular matter in the silo neck occurs at constant density approximately, but the flow lines are converging in the neck so that preservation of flow imposes that $v_g z^{d-1}$ is independent of z, so that speed $v_g$ of grains increases as $v_g \propto v_{go} (z_o/z)^{d-1}$ when approaching the neck, *i.e.* when z→B

D , where B is a constant and D the diameter of the opening. It means that the flow is accelerating near the neck, and the acceleration obeys $dv_g/dt = dv_g/dz \, dz/dt = v_g \, dv_g/dz$. As grain acceleration can not be larger than g in free fall this imposes a maximum speed and a hydraulic jump in the case of a permanent flow. This leads to a typical speed $v_s$ at the outlet which scales as

$$V_s = (5dg)^{1/2} = (Dg)^{1/2} \tag{5}$$

Let us now consider the restart of a flow after some stop; there shall be a transient regime lasting some lapse of time $\Delta t'$, during which grains are accelerated by gravity before they reach the permanent regime characterised by Eq. (5) at the outlet. During this transient, grains are accelerated by gravity. This imposes $V_s = g\Delta t'$ or $\Delta t' = (D/g)^{1/2}$. So using Eq.(5) and last relation fixes $\Delta t'$. During $\Delta t'$ particles have been uniformly accelerated, so that they have travelled from a distance $\xi = \frac{1}{2} g \Delta t'^2 = \frac{1}{2}D$ before the outlet, so that $\xi = \frac{1}{2}D$. Hence the typical flow which is concerned by the transient behaviour (before the permanent regime occurs) occupies a volume $\Omega_c = \pi D^2 \xi / 4 = \pi D^3/8 = 50 d^3$, since D=5d. It is tempting to compare this value to the volume of granular medium in the avalanche mode, *i.e.* $\pi d^3 s^*/[6(1-\varphi)] = d^3 s^*$ where $\varphi=0.5$ is the mean porosity. As $s^* < 50$ in experiments far from $R=R_c$, it means that stationary regime is only merely fulfilled when the avalanche size reaches the avalanche mode size $s^*$. On the contrary, as $s^* \approx 50$ in experiments at $R=R_c$ one gets the avalanche mode size $s^*$ to be equal to the transient regime.

***Seventhly,*** as it was argued in point (iv), correlations of fluctuations shall/can occur before stoppage of the silo in a different way than in the permanent regime. This will concern a short time $\Delta t$ before the stoppage. It is then expected that the correlation function $<Q(t)Q(t-\tau)>$ should vary with $\tau$ for $0<\tau < 10 \, N_c/<Q>$. A question arises: is it possible to study the correlation of flow fluctuation and under which accuracy and conditions? In practice, this shall mean that one shall be able to measure the typical number $N_c$ of grains which flow with time. Be $v_s$ the typical speed of a grain at the outlet; $V_s = (5dg)^{1/2} = (Dg)^{1/2}$ so that the typical time $\delta t$ for a grain to go out is $\delta t = d/v_s$.

To be able to follow the flow with a scale (as it is done in [1]) requires a flow rate of 1 grain/s about, or $V_s/d=1$ which requires grains/blocks larger than 10m.

However one could follow the flow and analyse it via optical detection of grain at the outlet: For 1mm grain, one gets $V_s/d = (5g/d)^{1/2} = 200$ and $N_q = (5^{5/2}/4)\pi(g/d)^{1/2}$ grains/s = $44 \, (g/d)^{1/2}$ grains/s = 4400 grains/s. This is feasible likely.

***Eighthly,*** as mentioned already the statistics of flow size s is described by a single Poisson process. It means that the blockage cannot be the conjunction of two or more independent events very delayed from one another, and occurring at any time during the flow otherwise the statistic would be modified. However, this statistic can be generated by a set of N events occurring merely at the same time.

*Ninthly,* in order to investigate such a possibility it might be interesting to investigate the *jamming with 2 kinds of balls* : let us assume (i) that jamming needs the formation of a vault or of a part of vault with a definite structure made of N grains, (ii) that the silo is filled with 1-x grains of size d and x grains of smaller grains (size d'<d) and (iii) that the flow does not stop when the local structure contains a smaller grain d' when the system is very near jamming. Then the new probability of stopping is $(1-x)^N P_N$ , where $P_N$ is the stopping probability with all grains being identical with size d. Then the measure of the new probability distribution may allow to determine N.

The angle of the opening may also plays some part : be $2\alpha$ the angle of the cone of converging flow line (in general $2\alpha=2*60°$ about). The number of grains on the surface of the blockage vault is $N=2\pi [1-\cos(\alpha)]R_c^2$ about in 3d (which reads $N_{3d}=\pi R_c^2$ =150grains with $\alpha=60°$), or $N_{2d}=2\alpha R_c$ (=10grains) about in 2d. The determination of the distribution of flow size as a function of the aperture angle $\alpha$ may allow determining whether the blockage structure is some small part of the whole vault or the whole vault itself, or if it grows as a volume. (If the vault formation requires some extension into the silo the number of grains involved in its formation shall be larger than the one which has been just computed; this shall change the statistics of avalanches and their mean typical size.)

Since $N_{3d}$ is expected to be 150 about (from last paragraph), it seems that 0.1%-1%-10% of smaller or larger grains introduced in the system should change the statistics and/or the threshold of jamming in 3d.

An other question is the change of the threshold $R'_c$ of the maximum diameter ratio when a mixture of two grains is used: Is $R'_c =R_c$, where $R_c$ is the threshold ratio obtained with the larger grains d, or does $R'_c$ depend on the composition x and on D/d' as $R'_c =R'_c(x,d/d')$. Is $R'_c(d>d')-R_c= R_c -R'_c(d<d')$ ? These are an important questions whose answer may help understanding the mechanics of vault jamming.

**As a conclusion,** in this simple well-controlled experiment reported in [1], the blockage of the silo looks controlled by a single Poisson process with no time correlation except the time needed to reach stationary flow. This single process is the "complete silo blockage", *i.e.* the formation of a stopper; everything looks as if the formation of this stopper occurs randomly with a probability 1-p at the passage of each grain. 1-p is found to depend on the distance $R_c$–R of the ratio R=D/d of the diameter of the silo outlet to the grain diameter. Here $R_c$ is the critical ratio above which no stopping occurs anymore.

It is most likely that $R_c$ itself depends on different parameters such as humidity and vibration, dispersion of grain size…. , so that the system may look more complicated in a less controlled experiment. For instance, letting the system evolve at different humidity w, one shall probably define a critical radius $R_c(w)$ which epends on moisture w. But the flow of the system will still be controlled by what occurs at the outlet, so that the hysteresis will look small (*i.e.* different from the hysteresis of the avalanche process).

However, it may also exist some experimental protocol which leads the silo flow to be controlled by what happens in the whole silo. In such a case one should observe a hysteresis whose size scales as the content in the silo. In this case the rheological law of the material will be important [7] and the silo structure also, which can favour blockage.

*Acknowledgements:* CNES is thanked for partial funding.